# Oxide-Semiconductor Materials for Quantum Computation

Jeremy Levy

*Center for Oxide-Semiconductor Materials for Quantum Computation*

*Department of Physics and Astronomy, University of Pittsburgh*

*Pittsburgh, PA  15260  USA.  E-mail: jlevy@mailaps.org*

PACS numbers:  03.67.Lx, 73.21.La, 77.84.2s

**ABSTRACT**

Scalable fault-tolerant quantum computer architectures require quantum gates that operate within a small fraction of the qubit decoherence time and with high accuracy over a bandwidth set by the decoherence rate.  Electron spin quantum bits in Si are promising because of long decoherence times (~0.5 ms), but electrical gating schemes still seem problematic.   Oxide-semiconductor heterostructures have the potential for precise electrical control of gate operations in the semiconductor, using optical rectification in the ferroelectric oxide.  Accurate (~80 dB), local (~10 nm), dynamic (>THz) and programmable optical control over electric polarization in the ferroelectric can be achieved using existing technology.  Optical techniques may also be useful in rapid initialization of the quantum computer, and for providing a source of initialized qubits to use for quantum error correction.  Advantages of optical methods will be discussed within a framework proposed for a quantum information processor using ferroelectrically coupled electron spins and Ge quantum dots in Si.

## Introduction

The physical requirements for scalable quantum computation are daunting. Stripped of their exceptional power, *e.g.*, quantum algorithms with exponential[1] or quadratic[2] speedups over their classical counterparts, quantum computers would hardly be worth building. One of the experimental challenges in constructing a quantum computer is the management of decoherence, the unavoidable entanglement between physical quantum bits and the environment. Optimization of decoherence times is essentially a materials issue, and the identification and quantification of decoherence an important theoretical and experimental activity.

A second experimental challenge relates to the "analog" character of quantum logic. While quantum computers process qubits just as classical computers process bits, quantum digital logic (one-qubit and two-qubit gate operations) more closely resembles that of analog computers. That is, both the gate strength and the time are essentially continuous parameters that must be controlled with a high degree of accuracy. The theory of fault-tolerant error correction[3] has shown that both decoherence and gate inaccuracies can be managed, provided they are sufficiently small. This error threshold $\varepsilon = \varepsilon(A) \sim 10^{-4} - 10^{-5}$ is believed to depend on the system architecture $A$, being for example smaller for architectures with only nearest-neighbor two-qubit gates.

The interplay between the constraints related to decoherence times and gate accuracy has some subtle aspects. For example, Heisenberg exchange between electron spins $\vec{s}_1$ and $\vec{s}_2$ (and the associated square-root-of-swap gate $SWAP^{1/2}$) is universal when combined with single spin rotations[4]. The $SWAP^{1/2}$ gate can be characterized by an exchange strength $J$ and a time $T$:

Eq. 1 $$SWAP^{1/2} = \exp\left[-i\int_0^T J\,\vec{s}_1 \cdot \vec{s}_2 / 2\hbar\right]$$

Assuming for convenience that a constant pulse $J(t) = J$ is applied for a time $T$, the constraint becomes: $J\,T = constant$. Fault-tolerant error correction requires that $\delta(J\,T)/J\,T = \delta J/J + \delta T/T < \varepsilon$. Decoherence of spin quantum information occurs within a time $T_2$, leading to the constraint $T < \varepsilon\,T_2$. Similarly, $J$ must be controllable over a bandwidth BW > 1/T, with a signal-to-noise ratio (SNR), integrated over BW, that exceeds $\varepsilon^{-1}$.

To meet the physical requirements for quantum computation, one must therefore choose qubits and qugates with sufficiently high speed, high accuracy, and low decoherence. While the gate operation time must be small relative to the decoherence time, the decoherence time alone provides severe constraints on the accuracy required for the control parameter $J$. Generally, the control parameter $J$ is controlled by a classical field (*e.g.,* electric field $E$). This additional functional dependence leads to an additional factor of $|dJ/dE|$, which may be reduced through careful design[5].

## Comparison Between Optical and Electrical Gating of Spin

Optical techniques are attractive for spin-based quantum computation due to the speed and accuracy with which they can be controlled. A value of $T_2$~0.5 ms requires that quantum gates operate within a time $T < 5\,\text{ns}$, assuming $\varepsilon = 10^{-5}$. Friesen *et al.* have surveyed existing electrical pulse generators, which have been proposed for electrical gating of Si-based quantum computers[5]. They find that $\delta V/V \sim 10^{-2}$ for the best available GHz-range pulse generators, which fall well below the requirements for quantum computation (assuming that $|dJ/dE| \sim 1$). In addition, jitter in the pulse length is $\delta T \sim 100\,\text{ps}$, leading to the requirement that $T > 10\,\mu\text{s}$, which is incompatible with the requirement set by decoherence. So, for existing electrical pulse technology to work for electrically gated electron spin-based quantum computing (assuming $T_2 \sim 0.5\,\text{ms}$), combined improvements to either pulse jitter and/or amplitude stability must exceed approximately $10^3$.

Optical fields are well suited to meeting the bandwidth and accuracy requirements of quantum gates. It was described above how optical fields can modulate polarization at Terahertz bandwidths. What are the noise specifications of the optical sources that create these polarizations? Femtosecond lasers have remarkable performance in terms of bandwidth, stability and timing accuracy. The lasers themselves can be exceedingly quiet, approaching the shot-noise limit. For example, the amplitude and timing noise of an externally mode-locked femtosecond laser operating at 10 GHz repetition rate was recently measured by Yilmaz *et al.*[12]. They found that an intensity noise of 0.05%, from 10 Hz - 5 GHz, less than 2dB above the shot-noise limit. The timing jitter over the same bandwidth was 240 fs. While the timing accuracy falls well within the requirements for quantum computation with electron spins in Si the amplitude stability can be improved by increasing the number of photons per pulse, or by quantum squeezing of the amplitude noise. Optical sources have other spectacular features that will not be discussed in to great detail here. They include: (1) attosecond control over the optical phase that can be used for coherent control of electronic degrees of freedom[13]; (2) metrological "lock-to-clock" accuracies[14] better than a part in $10^{-15}$, (3) natural gateways between stationary and "flying" qubits that can be used for quantum key distribution[15].

## Oxide-Semiconductor Materials

Within these constraints, what is the best choice of quantum bit and quantum gate? The answer depends on many materials and technological issues. This paper will discuss one particular approach to the problem, and will try to motivate the various choices that have been made in the context of the above discussion.

Table 1 lists the requirements for scalable quantum computation[6], along with a summary of the approach discussed here[7]. Requirement (R1) is addressed by using electron spin qubits in Si. The motivations for using spin[4] and Silicon[8] are well discussed in the literature. Requirement (R2) is addressed by using circularly polarized light to create spin-polarized electrons in Si using small Ge quantum dots, a method that has been used very successfully in zincblende semiconductors[9]. Strained Ge has type-II band offsets with Si, so a photoexcited electron-hole pair will spatially separate, with the hole tightly confined in the Ge, and the electron loosely bound to the dot by the electrostatic interaction with the charged Ge dot. Requirement (R3) is addressed by a combination of long intrinsic decoherence times for Si ($T_2 \sim 0.5$ ms [10]) and fast electrical gates (~THz) that work from a combination of static and dynamic polarization from an epitaxial ferroelectric thin film.

The basic architecture of the proposed quantum information processor is illustrated in Figure 1. Ge quantum dots are grown in a controlled linear array on the surface of Si. The dots are spaced such that each can be optically addressed. An epitaxial ferroelectric oxide is grown on top of the semiconductor. The requirements for the ferroelectric are that it be uniaxial with a spontaneous polarization pointing perpendicular to the plane. In addition, there must be a sufficiently low density of interface traps so that the static polarization of the ferroelectric can have a field effect in the semiconductor. The top view, shown in Figure 1(b) illustrates how the ferroelectric can be polarized (using a conducting atomic force microscope probe) to create confined regions for electrons.

In addition to providing static confinement for electrons and their spin, the ferroelectric plays an essential role in gating electron spin interactions mediated by Heisenberg exchange. Figure 2 illustrates how optical rectification can modulate the static polarization in the ferroelectric. Figure 2(a) shows a situation in which two electrons reside in a potential well defined by the static ferroelectric polarization. A small barrier (~10 nm) separates the ferroelectrically-defined quantum dots. Upon uniform illumination of the ferroelectric, the *magnitude* of the polarization is decreased, resulting in a lower potential barrier for the two electrons. Resulting overlap of electron orbitals leads to well-known Heisenberg exchange between the spin degrees of freedom. When the light is turned off again, the barrier returns to its former height.

A few remarks should be made about the optical rectification process. First, the wavelength of light can be chosen so that direct absorption in the semiconductor or ferroelectric do not occur. Otherwise, the introduction of free carriers would inevitably disrupt the system. Second, the optical resolution required for controlling the potential barrier for exchange is limited not by diffraction, but by the ability to write domains in thin ferroelectric films. Finally, it should be noted that similar gating schemes based on optical Stark shifts has already been applied successfully for the optical manipulation of electron spins in III-V material on femtosecond time scales[11].

Support for this work by DARPA QuIST (DAAD19-01-1-0650) is gratefully acknowledged.

# Cited References

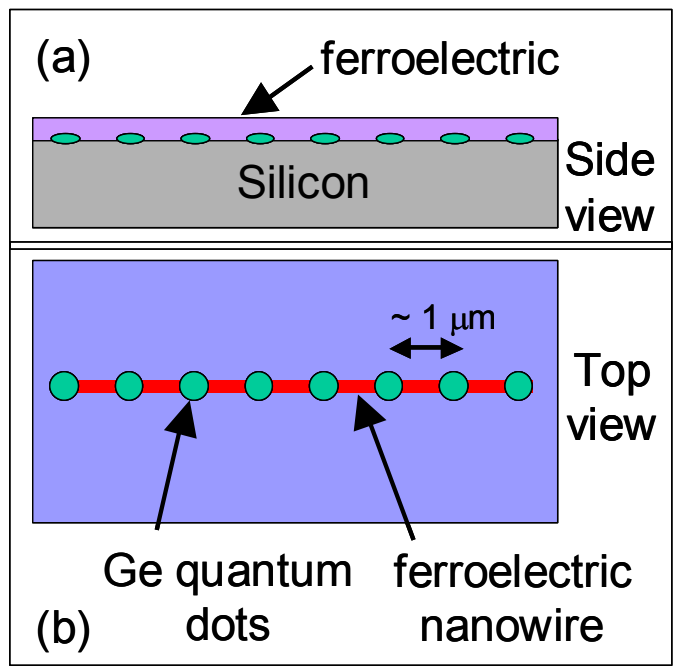

Figure 1

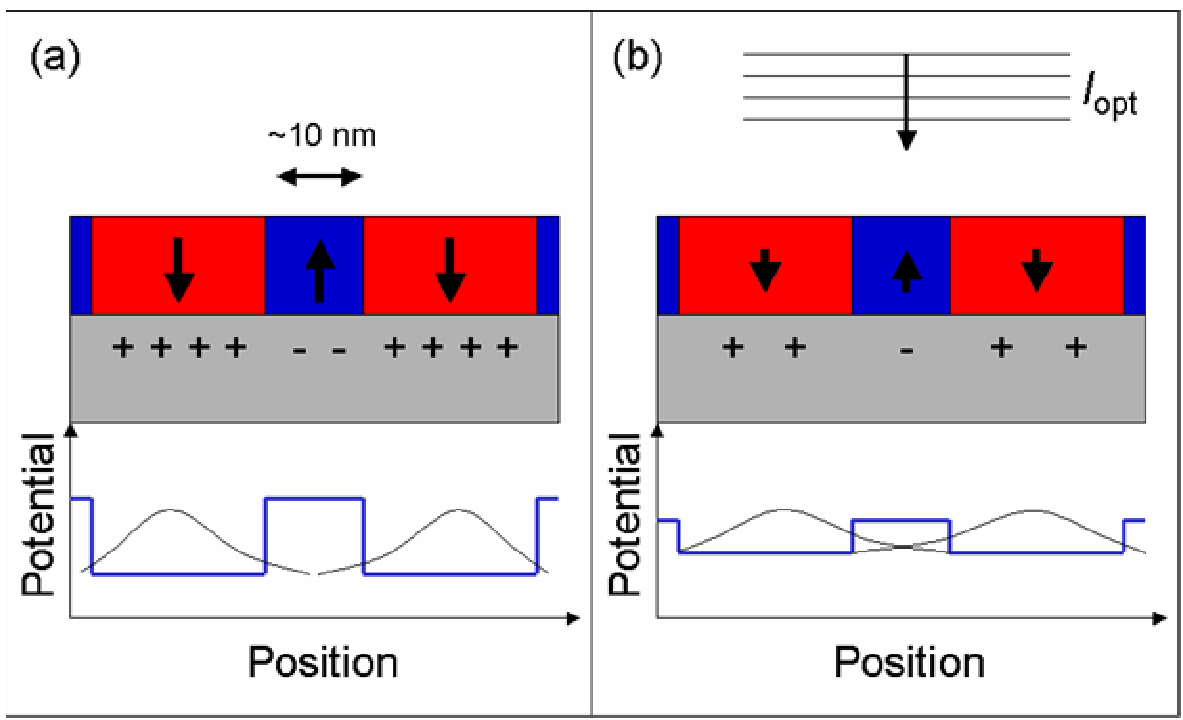

Figure 2

Table 1

| | **Requirement** | **Proposed Implementation** |
|---|---|---|
| (R1) | A scalable physical system with well-characterized qubits | Electron spins in Si |
| (R2) | The ability to initialize the state of the qubits to a simple fiducial state, such as \|000…0> | Optical spin injection into Si using quasi-direct gap Ge quantum dots |
| (R3) | Long relevant decoherence times, much longer than the gate operation time | Long spin lifetimes in Si; fast (2-qubit) gate operation times using ferroelectric gates |
| (R4) | A "universal" set of quantum gates | g-factor engineering and Heisenberg exchange using optical rectification in ferroelectric |
| (R5) | The ability to measure specific qubits | Single electron transistors (non-optical) |

Table 1

Figure Captions

Figure 1. Schematic of proposed quantum information processor. (a) Ge quantum dots "sandwiched" between Si substrate and epitaxial oxide ferroelectric thin film. (b) Ferroelectric polarization can be patterned by a conducting scanning probe to create nanostructures in the semiconductor.

Figure 2. Schematic illustrating the effect of optical rectification. (a) Two electrons are confined by ferroelectric potential. (a) Application of a sub-gap optical field reduces the magnitude of the ferroelectric polarization, and increases the electron orbital overlap, leading to Heisenberg exchange.

Table Captions

Table 1.

Requirements for scalable quantum computation, and proposed method for meeting these requirements.